\documentclass[sigconf,10pt]{acmart}
\usepackage{enumitem} % compact bullet points
\setlist{topsep=5pt, leftmargin=*}

\usepackage{siunitx} % write Units in correct way e.g \SI{100}{\meter\per\second}  % Displays "100 m/s"
\AtBeginDocument{%
  \providecommand\BibTeX{{%
    \normalfont B\kern-0.5em{\scshape i\kern-0.25em b}\kern-0.8em\TeX}}}

\setcopyright{acmlicensed}
\copyrightyear{2024}
\acmYear{2024}
\acmConference[NetAISys '24]{International Workshop on Networked AI Systems}{June 07,
  2024}{Tokyo, Japan}
\copyrightyear{2024}
\acmYear{2024}
\setcopyright{licensedothergov}\acmConference[NetAISys '24]{The 2nd Workshop on Networked AI Systems}{June 3--7, 2024}{Minato-ku, Tokyo, Japan}
\acmBooktitle{The 2nd Workshop on Networked AI Systems (NetAISys '24), June 3--7, 2024, Minato-ku, Tokyo, Japan}
\acmDOI{10.1145/3662004.3663554}
\acmISBN{979-8-4007-0661-5/24/06}

\begin{document}

% \title{Enhancing UWB for enabling sovereign data network in hospitals}
\title{Advancements in UWB: Paving the Way for Sovereign Data Networks in Healthcare Facilities }

\author{Khan Reaz}

\orcid{0000-0001-9857-9687}
\author{Thibaud Ardoin}
\author{Lea Muth}
\author{Marian Margraf}
\author{Gerhard Wunder}
\orcid{0009-0001-0850-8816}

\affiliation{%
  \institution{Freie Universität Berlin}
  % \streetaddress{P.O. Box 1212}
  \city{Berlin}
  \country{Germany}
  \postcode{14195}
}

\author{Mahsa Kholghi}
\author{Kai Jansen}
\author{Christian Zenger}
\affiliation{%
  \institution{PHYSEC~GmbH}
  \city{Bochum}
  \country{Germany}}

\author{Julian Schmidt}
\author{Enrico Köppe}
\affiliation{%
  \institution{NC Systems~GmbH}
  \city{Berlin}
  \country{Germany}
}

\author{Zoran Utkovski}
\author{Igor Bjelakovic}
\author{Mathis Schmieder}
\affiliation{%
 \institution{Fraunhofer Heinrich-Hertz Institut}
 \city{Berlin}
 \country{Germany}}

\author{Olaf Dressel}
\affiliation{%
  \institution{Bundesdruckerei GmbH}
  \city{Berlin}
  \country{Germany}}

% \affiliation{%

% \affiliation{%

% \affiliation{%

\renewcommand{\shortauthors}{Reaz and Ardoin, et al.}

\begin{abstract}
Ultra-Wideband (UWB) technology re-emerges as a groundbreaking ranging technology with its precise micro-location capabilities and robustness.  This  paper highlights the security dimensions of UWB technology, focusing in particular on the intricacies of device fingerprinting for authentication, examined through the lens of state-of-the-art deep learning techniques. Furthermore, we explore various potential enhancements to the UWB standard that could realize a sovereign UWB data network. We argue that UWB data communication holds significant potential in healthcare and ultra-secure environments, where the use of the common unlicensed 2.4~GHz band-centric wireless technology is limited or prohibited. 
A sovereign UWB network could serve as an alternative, providing secure localization and short-range data communication in such environments.

\end{abstract}

\copyrightyear{2024}
% \conferenceinfo{NetAISys '24,}{June 3--7, 2024, Minato-ku, Tokyo, Japan}
% \isbn{979-8-4007-0661-5/24/06}
% \doi{10.}



% \ccsdesc[500]{Security and privacy~Security protocols}





\maketitle

\section{Introduction}
The unprecedented resurgence of Ultra-Wideband~(UWB)  technology in consumer electronic devices is attributed to its unparalleled accuracy in indoor tracking, minimal interference with other wireless technologies, and low cost. As outlined by the FiRa Consortium~\cite{FiRausecase}, UWB possesses huge potential to become the de facto standard for providing precision secure ranging and proximity authentication. Among its many potential use cases, UWB’s distinctive features are particularly well-suited to the requirements of hospital settings, focusing on ensuring patient safety, operational efficiency, and the dependability of medical equipment and communications. One exemplary application is asset tracking and access control. Keeping track of medical equipment, such as portable monitors, infusion pumps, wheelchairs, and other valuable assets, is crucial in hospitals or nursing homes. In this system, UWB tags are affixed to different pieces of equipment. These tags communicate with UWB anchors installed throughout the facility, providing precise location data for each tagged item. This information is then integrated into the hospital’s digital management system. Beyond asset tracking and access control, the data exchange capability of UWB technology presents an attractive opportunity. This involves interconnecting potential equipment in a peer-to-peer and peer-to-mesh topology, allowing them to exchange small amount of data.  Standards such as the one from FiRa, Car Connectivity Consortium~(CCC), and Omlox primarily consider the secure ranging aspect of the UWB technology but leave out its data communication capabilities. Much of the current UWB-related literature focuses primarily on achieving greater ranging accuracy, faster refresh rates, and improved efficiency for mobile tags. The data communication aspects of the UWB nodes, however, remain less explored or simply ignored. From our perspective, data communication over the UWB interface has potential in healthcare sectors, special industrial facilities, and ultra-secure governmental premises, where the security guarantee provided by UWB technology may outweigh its energy aspects. In these locations, the use of common unlicensed 2.4/Sub-GHz band-centric wireless technology is restrictive. UWB could serve as an alternative, offering not only secure localization but also resilient  data communication in such environments.

This paper closely examines the current state of UWB technology, and identify gaps in the existing standards. In Sec.~\ref{sec:sec-enhancements}, we suggest several enhancements to make UWB more secure and practical. For instance, we propose integrating deep learning techniques for device authentication using physical layer information. This approach reduces the need for out-of-band or certificate-based authentication. Additionally, we have introduced the concept of \emph{Secrecy Maps}, a tool that makes it possible to determine the data leakage quantitatively in a statistical manner and visualize it spatially under different channel conditions.  Furthermore, in Sec.~\ref{sec:data-network}, we propose a UWB network architecture designed specifically for healthcare facilities and other similar environments. This architecture is further enhanced by a new Medium Access Control (MAC) protocol that enables UWB devices to communicate within a mesh network.  Such a network is more flexible, resilient, and scalable.  Sec.~\ref{sec:implementation} presents the implementation details. Finally, we summarize our findings and recommendations in Sec.~\ref{sec:conclusion}. Altogether, our proposed enhancements are aimed to solidify UWB as a secure, and reliable technology. 

\section{UWB Security Enhancements}
\label{sec:sec-enhancements}
 \subsection{Machine Learning Based Fingerprinting}

    % ** What is fingerprint
    Radio Frequency Fingerprinting~(RFF) is a type of signal intelligence applied within the radio frequency domain.  It involves  a technique that identifies a unique signature from the hardware  transmitting the signal~\cite{Xie2023}. This unique identification  results from the unintentional variations introduced during the fabrication process of different physical components resulting in slight changes in the signal's waveform without altering the intended transmitted data\cite{reaz:compass}. 
    % Precisely, we use the following definition of the fingerprint~\cite{Xie2023}:%
    % \begin{itemize}
    %     \setlength\itemsep{-0.1em}
    %     \item \emph{Differentiable}: Each device has a unique fingerprint that can be distinguished from those of other devices.
    %     \item \emph{Relative stability}: The unique feature should remain  as stable  as possible over time, despite environmental changes.
    %     \item \emph{Hardware}: The sole independent source of the fingerprint is the condition of the hardware. Any other impact on the waveform, such as interference, temperature, time, position, orientation, or implementation, is considered a bias.
    % \end{itemize}

    % ** How it gets used
    Once RFF features are extracted from the signal, they can be used as a unique identifier for a particular device. Since this fingerprint is rooted in the hardware, and derived from the raw signal shape, it is inherently challenging to mask or spoof. Therefore, RFF has the potential to become a key element of physical layer security, offering a robust method for \emph{Device Authentication}.  

    % ** Specific to UWB
    Most studies demonstrating successful device classification via RFF focus on commonly utilized radio technologies, such as \mbox{Wi-Fi}, 5G, or Bluetooth~\cite{Jagannath2022}. To the best of our knowledge, there has been no research conducted on RFF for UWB signals thus far. 
    Two technical aspects could add challenges to the RFF detection in the case of UWB signal.
    First, UWB technology transmits pulses, resulting in shorter duty cycles compared to the continuous transmission seen in other technologies. This leads to fewer available features in the signal for fingerprint detection. 
    Second, the main advantage of UWB technology is its position sensitivity, rendering it advantageous for numerous applications. Nevertheless, this characteristic can introduce strong variations in the signal when the position changes, which ultimately hampers the learning of consistent features across diverse environments or locations. 
    
    % Figure~\ref{fig:dev-position-comparison}  illustrates how changing the position of one device has a more significant impact on the signal than swapping devices.

    % \begin{figure}[h]
    %     \centering
    %     \includegraphics[width=\columnwidth,trim={0 0 1.5cm 0},clip]{Source/Graphics/signal_same_device.png}%
    %     \caption{Signal amplitude shape according to position and device.}%
    %     \label{fig:dev-position-comparison}
    % \end{figure}

    % \begin{figure}
    %     \centering
    %     \includegraphics[width=\columnwidth,trim={0 0 1.5cm 0},clip]{Source/Graphics/signal_same_position.png}
    %     \caption{Caption}
    %     \label{fig:enter-label}
    % \end{figure}

% \subsection*{Data Acquisition} 

    % Addressing potential measurement biases is crucial, despite the theoretical concept of fingerprint. Unintended biases may lead machine learning models to differentiate devices based on incidental factors rather than genuine hardware signatures — a phenomenon referred to as the \emph{Clever Hans effect}. 
    Unfortunately, current scientific literature on RFF often overlooks potential measurement biases, with only a few works presenting compelling evidence of mitigating biased correlations associated with detection. 
    In order to minimize the bias due to position change during the data acquisition process, we utilized a  3D printed mount. The Channel Impulse Response~(CIR)  of the signal is then measured with two UWB development boards (Qorvo's 3001CDK) —the receiver and the emitter— in a fixed position facing each other. The mount is then gradually rotated in a clockwise manner to generate CIR data across different positions in the environment. 
    
    We consider the following three evaluation cases to clearly demonstrate how well the learning process can generalize its feature extraction:

\begin{itemize}
    \setlength\itemsep{-0.1em}
    \item \emph{Case 1: } The fingerprint authentication is conducted between two fixed nodes (typically known as \emph{anchors}) within a UWB network, where both the positions and identities of the nodes are known. The constructed model is fine-tuned based on this information.

    \item \emph{Case 2: } The fingerprint authentication is performed between a fixed node and a \emph{tag} node moving around the environment. The tag's identity is known, and the feature extraction is fine-tuned specially for this tag.

    \item \emph{Case 3: } The fingerprint authentication is carried out between an anchor node and an unknown tag in an unknown position. Initially, the network generates a trusted feature embedding of that tag to serve as an ID. Future feature embeddings can then be compared to this anchor to re-identify the tag.

\end{itemize}

\subsubsection{Specialized Deep Learning Approaches}

    % WHY DIFFERENT APPROACHES
    Ideally, algorithms designed for~RFF should be lightweight enough to facilitate real-time identification. However, achieving accurate RFF necessitates striking a balance between computational efficiency, and the algorithm's overall performance. Here, we introduce two approaches that address the trade-offs along this spectrum:

    % \emph{\textbf{Convolution Neural Network:}} 
    A classical method highlighted in the literature~\cite{Ding2018, AlShawabka2020} for dealing with signal data and identifying different sources in RFF is the use of 1-dimensional Convolution Neural Network (CNN) to predict the source device. 
    With this specific feature extraction, we can build a small classifier model of 10k parameters that can run on a low resourced device.
    The performances of such a CNN model can be found in Table \ref{table:ML results}. Noticeably, in Case 1, the results have 100\% success in classification. However, when we introduce positional variation, as seen in Case 2 and 3, the performances come closer to random. This indicates an over-fitting to the positional dependency of the training data and poor ability to generalize across different environments.
    
% \subsubsection{Transformer}
    In order to extract more potent features from the signal, we opted for a vision Transformer (ViT)~\cite{Vaswani2017, Dosovitskiy2021}, using spectrograms of the signal as input. This model has been trained using a contrastive learning method~\cite{Yuan2019}, a paradigm of deep-learning that project the data into a high-dimensional latent space to capture only the relevant details necessary for the problem. Inspired by the work performed on biometrical identification such as facial recognition~\cite{schroff2015}, this method is well-suited to the re-identification problem. The technique aims to achieve more generalization as the model is explicitly trained to ignore the positional information. In this setup we managed to improve the results for the second and third evaluation scenarios, indicating a better generalization of the desired feature extraction.

    %    \begin{table}[htbp]
    %      \centering
    %      \caption{Performances for RFF. The reported metrics are F1-score for Classification and F1-score for the Re-Identification task.}
    %      \begin{tabular}{lllllll}
    %      %\toprule
    %      & CNN & &
    %      ViT &  &
    %      Random \\
    %      \midrule
    %      & Class & Re-ID & Class & Re-ID & Class & Re-ID \\
    %      \midrule
    %      Case 1 & $\mathbf{100\%}$ & $77\%$ & $100\%$ & $\mathbf{100\%}$ & $9\%$ & $9\%$ \\
    %      Case 2 & $15\%$ & $26\%$ & $39\%$ & $\mathbf{56\%}$ & $10\%$ & $10\%$ \\
    %      Case 3 & $11\%$ & $23\%$ & $33\%$ & $\mathbf{42\%}$ & $8\%$ & $9\%$ \\
    %      \bottomrule
    %      \end{tabular}
    %      \label{table:ML results}
    %      \end{table}

       \begin{table}[htbp]
         \centering
         \caption{Performances for RFF. The reported metrics are F1-score for Classification and F1-score for the Re-Identification task.}
         \begin{tabular}{cccrrr}
         %\toprule
        & & & Case 1 & Case 2 & Case 3 \\
         \midrule
        CNN & Classification & & $\mathbf{100\%}$ & $15\%$ & $11\%$ \\
        \textit{(10k)} & Re-ID & & $77\%$ & $26\%$ & $23\%$\\ 
         \midrule
        ViT & Classification & & $\mathbf{100\%}$ & $39\%$ & $33\%$ \\
        \textit{(2Mio)} & Re-ID & & $\mathbf{100\%}$ & $\mathbf{56\%}$ & $\mathbf{42\%}$ \\
         \midrule
        Random & Classification & & $9\%$ & $10\%$ & $8\%$ \\
         & Re-ID & & $9\%$ & $10\%$ & $9\%$ \\          
        \bottomrule
         \end{tabular}
         \label{table:ML results}
         \end{table}

\subsubsection{Results}
    The primary metric reported in our study is the F1-score for classification. It is a measure of the accuracy that a classifier can obtain in the latent space of the models. It displays the capacity of the model to shape a separable representations, but is not the method used in practice. Therefore, we introduce a second metric, the F1-score for re-identification. It consists of creating anchor vectors that represent each identity in the latent space that we compare to the query vectors to match the corresponding ID. This way we can also identify devices that have never been seen during the training process. 
    Note that these metrics are raw 1-shot result to give an idea of the performance. In practice, the accuracy could be improved by collecting multiple signal samples per device for identification purposes.
    The key insight gleaned from our work is the recognition of the challenges associated with implementing fingerprinting for UWB signals. 
    Simpler solutions, though viable, lack robustness and require fine-tuning in specific settings. Their effectiveness is significantly influenced by the stability of the environment, particularly in networks where fixed devices consistently interact with each other.
    In contrast, more sophisticated learning strategies, such as those we offer, provide improved device discrimination when data is collected under controlled conditions. 
    
    % Comparatively, the first open datasets for facial recognition such as CelebA~\cite{liu2015faceattributes} consists of approximately 10,000 different identities captured in various environmental conditions. However, our experiments contain recordings of only 13 unique devices in a limited variation of environment. Despite the imperfect results, this sample size suggest the potential for improved generalization in RFF if trained with data from more diverse environment.

% \end{itemize}

\subsection{Countermeasures to Secure ToA-based Ranging }

% \begin{figure}[H]
%     \centering
%     \includegraphics[width=\columnwidth]{Source/Graphics/toa-secure.png}
%     \caption{Countermeasures to protect   ranging through comparative analysis of ToA}
%     % \label{fig:enter-label}
% \end{figure}

UWB devices are vulnerable to Time of Arrival~(ToA) manipulation attacks in various ways. A robust countermeasure can be implemented by any node within the UWB system to enhance the security against ToA manipulation. This involves  using the UWB PHY synchronization header to accurately timestamp the ToA for critical messages such as the Ranging Initiation Message, Ranging Response Message, and Ranging Final Message. When a UWB node receives these messages, it conducts a comparative analysis between the ToA calculated from the STS and the ToA derived from the UWB PHY header. A significant discrepancy between these two ToA values can serve as a red flag, prompting the UWB node to invalidate the ToA and, consequently terminate the ranging process. 

\subsection{Spatial Characterization Through \emph{Secrecy Maps}}
In the wireless radio access scenario, the knowledge of the local geometry provides context information for the probabilistic secrecy characterization. In particular, 
% the probabilistic characterization necessitates prior knowledge of the distribution of large-scale channel gains between arbitrary locations in the radio access network. %, as integration is required over the set of all possible locations of the eavesdropper. 
second order statistics of channels between arbitrary locations and a fixed location in the radio access networks can be obtained from so-called radio maps, %which is a common way of representing this kind of information, and 
whose construction and estimation has been a popular research topic in machine learning in recent years~\cite{romero2016blind}. 

Based on this, we propose the use of any-to-any (A2A) radio maps that provide means for characterization of the wireless environment from the perspective of the adopted security metric (e.g., semantic security), by associating security levels to the communication links. This characterization results in so-called \textit{secrecy maps}~\cite{SPAWC} that may be considered as a generalization of the concept of radio maps for physical layer security. The specified security levels are based on mathematically provable security guarantees for certain channel models, with semantic security serving as the security metric~\cite{Goldwasser, Vardy, emara2018availability}. The flexibility of this approach has already been demonstrated by using it in the context of indoor and outdoor communication in traditional wireless networks \cite{SPAWC, Secrecy-Maps-Outdoor}, as well as in the case of sub-THz wireless communication \cite{schulz2021semantic}. In addition, variants of radio maps are already being discussed in the realm of UWB and used to solve specific localization problems (cf. for example \cite{Naghdi, Kolakowski} and references therein). Krentz et al. have identified achievable communication rates for certain UWB multipath fading channel models~\cite{SCREWED:2014, arikan2004capacity}. These results can serve as a starting point for construction of any-to-any radio maps which can then be augmented by security aspects. The basis for this will be a mathematical derivation of security guarantees for such types of channel models with respect to semantic security metric. 

\begin{figure}
    \centering
    \includegraphics[width=\columnwidth]{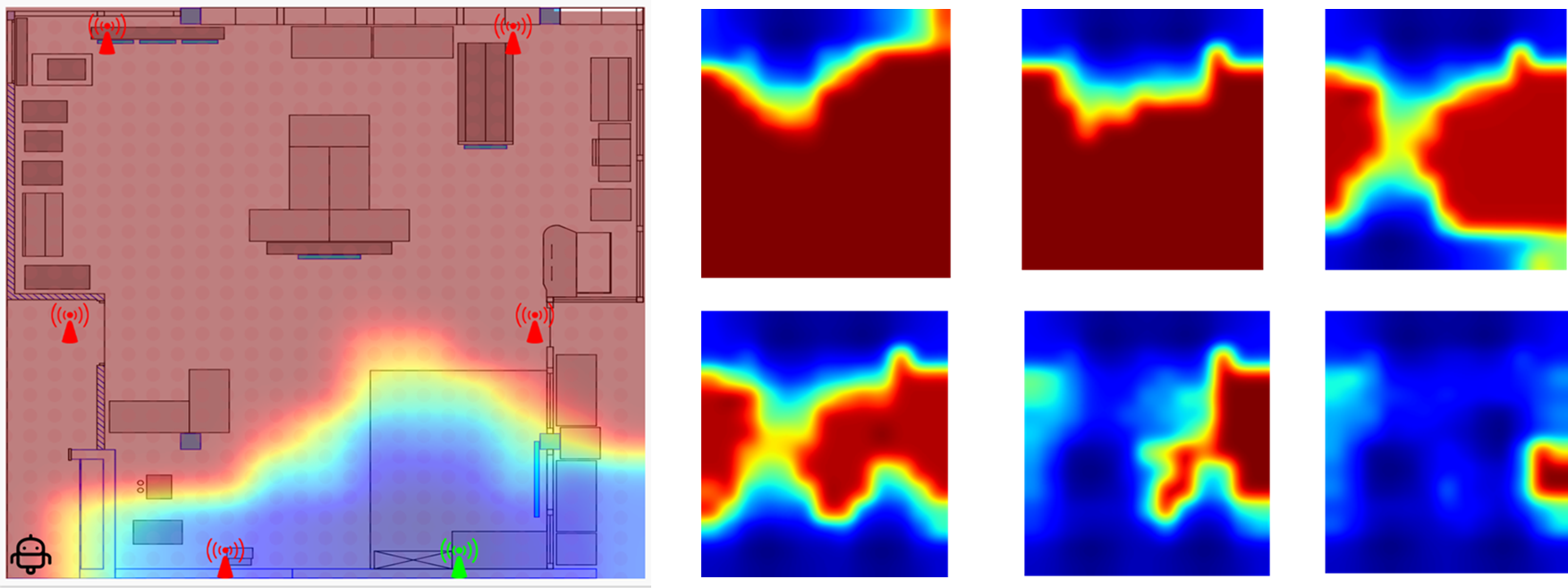}
    \caption{Qualitative depiction of a statistical secrecy characterization in an indoor environment (uplink scenario). Different colors encode different security levels. On the right-hand side we illustrate the effect of (gradually) switching on additional access points on the secrecy characterization (improvement of the secrecy outlook).}
    \label{fig:secrecy_map}
\end{figure}

The construction of secrecy maps can  be based on either ray tracing simulators, such as NVIDIA-Sionna, in which a virtual replica of the environment and the radio channel is created, or real channel measurements for specific spatial regions can provide the necessary data for the construction of secrecy maps. To illustrate the concept, in Fig.~\ref{fig:secrecy_map} we depict (qualitatively) a secrecy map generated based on measurements in an indoor scenario (uplink). For each different position of the transmitter, we plot the resulting secrecy level of the transmitter-receiver communication link, with semantic security serving as a metric. The secrecy characterization is performed in a statistical sense, meaning that the security level can be guaranteed with a pre-defined probability. 

\section{Sovereign UWB Data Network}
\label{sec:data-network}
\begin{figure}[ht]
    \centering
    \includegraphics[width=\columnwidth]{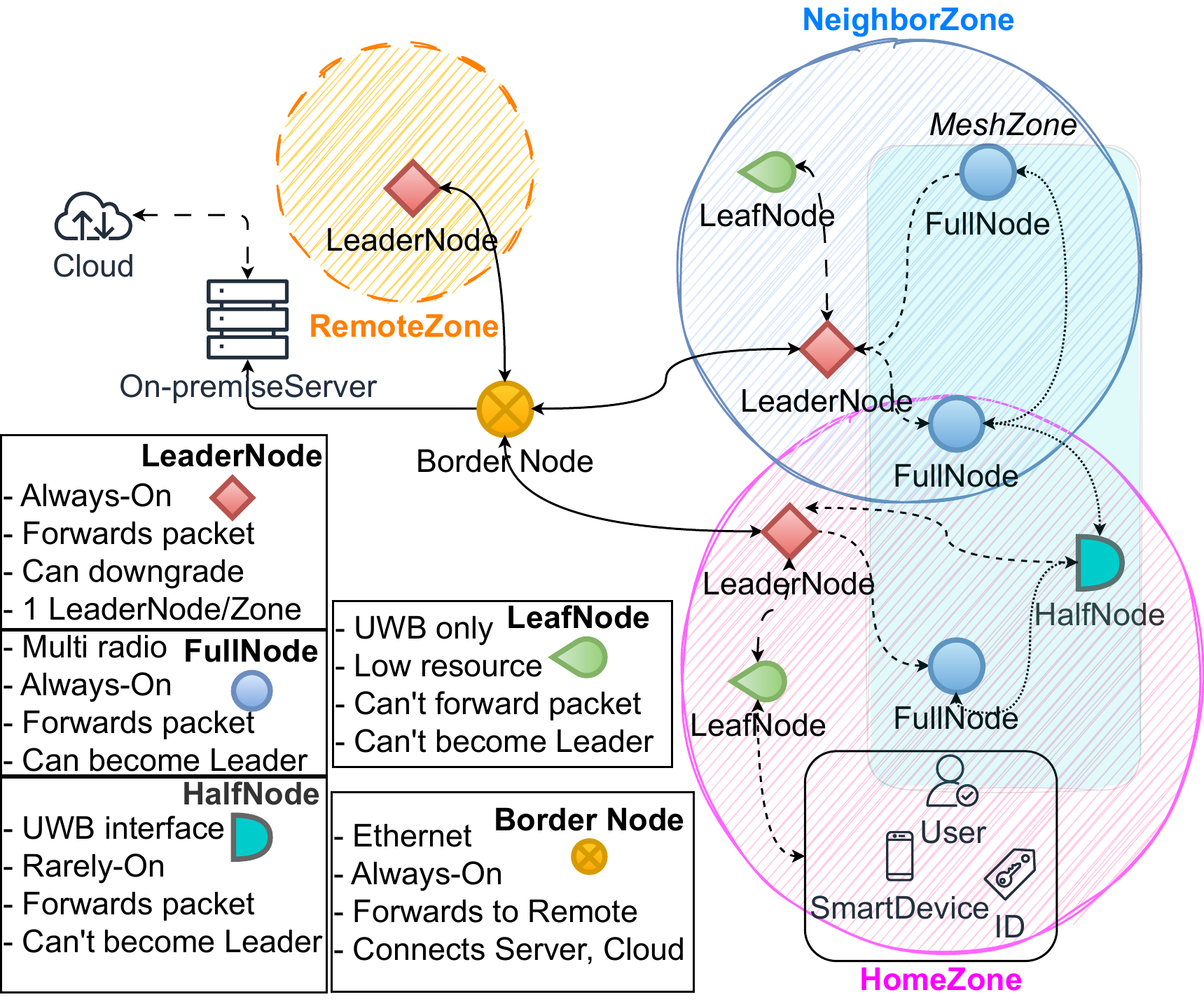}
    \caption{ Architecture of an ultra secure UWB data network}
    \label{fig:uwb-network-arch}
\end{figure}

\subsection{Network Architecture} We propose an architecture, depicted in Fig.~\ref{fig:uwb-network-arch} for a UWB data network, which diverges from the traditional \emph{anchor} and \emph{tags} concept by attributing devices based on their capability and availability.  \emph{LeaderNodes} are critical components that maintain constant operation; they are always-on and equipped with multiple network interfaces, which enable them to oversee packet forwarding and network synchronization. \emph{FullNodes} support the network with the added potential of being promoted to LeaderNode status. \emph{HalfNodes}, which are less frequently active, utilize UWB interfaces for packet forwarding but do not have the capacity to become a leader. \emph{LeafNodes}, with their limited resources, perform only ranging functions and do not forward packets or assume leadership roles. \emph{BorderNodes} stand out with their Ethernet connections, always active to ensure the connection between the mesh network and external zones such as the \emph{RemoteZone} and servers. We define the \emph{HomeZone} as the first point of contact for an Authenticated User and their Smart Devices with a UWB interface. During the bootstrapping phase, the system administrator of the facility setup an on-site server and defines key policy for each category of the  nodes. This key policy  contains attributes such as node to node distance constraints to enforce proximity bound, node's RFF  function, communication frequency (to thwart DDOS attacks).The users of the envisioned facility then use some government issued digital identity card to authenticate themselves in a privacy preserving way as proposed in~\cite{reaz:permission-voucher}.

\subsection{Medium Access Control Protocol} Even though the UWB PHY layer supports data communication, building a standalone UWB data network is challenging due to the absence of a standardized Medium Access Control~(MAC) layer protocol for applications beyond ranging. Typically, two approaches are taken to design a MAC protocol for wireless technologies: (1)~\emph{Spectrum sensing-based}, and (2)~\emph{Time-slot-based}. All commercially available wireless technologies are narrowband compared to the vast bandwidth present in UWB technology. The impulse-like nature of UWB signals (i.e., short duty cycle) poses challenges in implementing popular sensing-based MAC protocols, such as Carrier Sense Multiple Access with Collision Avoidance~(CSMA/CA), thus rendering them unsuitable for use.  In order to enable the UWB devices to form a data network, a UWB MAC stack has been developed. A key feature of this stack is the use of a \emph{superframe} structure. The superframe~(\SI{100}{\milli\second}) is divided into time slots using Time Division Multiple Access~(TDMA). This organized structure of the superframe ensures that devices have designated time slots for communication, reducing the likelihood of collisions. Each time slot serves a specific purpose and is allocated for a particular task. 

In a zone, a \emph{LeaderNode} periodically broadcasts superframes. Within one superframe, the \emph{LeaderNode} provides synchronization information, helping all devices in the zone maintain a common time reference. This synchronization is crucial for accurate ranging and coordination within the UWB network. The information within the beacons also provides the position and mesh topology of the corresponding \emph{LeaderNode}. Following the beacon slots, ranging slots are used by the other nodes for localization (e.g., through two-way ranging) and data exchange between a \emph{LeaderNode} and  another node. The last block of slots is reserved for data exchange between \emph{LeaderNodes} with each node having its own collision-free data exchange slot.

%%% rewrite to accomodate high-low density network.

% For this reason, we recommend that always-powered-nodes employ \emph{preamble hunting}-based channel access to avoid collisions within the radio zone. In this protocol, a node goes into receiving mode and samples the air for a short amount of time to look for a preamble from its predefined set (the UWB standard defines a set of available preambles for each UWB channel (e.g., Channel 5 has four preambles). For example, in systems based on the Qorvo DW3000, this is configured by \texttt{PRE\_TOC} to detect preamble. If the preamble is seen, \texttt{CCA\_FAIL} flag is raised and set back-off timer to one superframe to try again. If no preamble is detected, the node transmits. For battery powered nodes, an efficient power usage strategy is to transmit data along with a request for acknowledgment. This approach ensures that the node only expends energy on successful transmissions. If an acknowledgment is not received, the node activates its back-off timer, setting it to delay the next transmission attempt for the duration of one super-frame. This mechanism helps in conserving power by avoiding immediate retransmissions and optimizing the timing of subsequent attempts."

\subsection{UWB Mesh Networking}In our proposed UWB mesh network architecture, nodes are connected either directly to a  node or indirectly through other mesh nodes. This allows each node to communicate directly with its neighbors and   relay messages for other nodes, extending the network's reach (\emph{automatic routing}). The mesh network is resilient to node failures; if a node becomes unreachable, the network  dynamically reroute traffic through alternative paths (\emph{self-healing}). Adding more \emph{LeaderNodes/FullNodes} easily scales the mesh network; new nodes can join, and existing nodes will adapt the routing accordingly (\emph{scalability}). Smart devices with UWB capability  are handled differently in the UWB mesh network due to their mobility.  They primarily connect only during their ranging slot intervals (listening to beacon slots as required), operating in an energy-efficient manner. Consequently, mesh messages are exchanged within the ranging slots through the surrounding \emph{LeaderNode}.

\subsection{Payload Accommodation} 
IPv6 represents the latest version of the Internet Protocol~(IP), the core technology behind the Internet. 
% The transition from IPv4 to IPv6 is primarily driven by the need to address the limitations of IPv4, notably its limited address space, that is insufficient for the rapidly expanding number of devices connecting to the Internet. 
Beyond the expanded address space, IPv6 introduces enhancements in routing and network autoconfiguration, reduces the need for network address translation~(NAT), which can complicate Internet communication and hinder the performance of low-latency applications. IPv6 also includes improved security protocols as part of its core specifications, offering built-in support for IPsec for more secure communications. Clearly, UWB devices must be able to  handle IPv6 packets to remain as a future-proof technology. However, the direct integration of IPv6 packets, which are 1280 bytes in size, into UWB's 127-byte frames is impractical due to size constraints. The 6LoWPAN (IPv6 over Low-Power Wireless Personal Area Networks)  provides an essential  solution. One of its primary advantages is the ability to transmit IPv6 packets over low-power, low-bandwidth wireless networks, enabling devices with limited processing capabilities to connect directly to the Internet. Additionally, 6LoWPAN's efficient use of IP-based communication reduces the need for complex protocol translations, simplifying network architecture and lowering operational costs. Consequently, integrating a 6LoWPAN adaptation layer into the UWB standard is necessary to compress the 40-byte IPv6 header to a minimum of 3 bytes.

% The ever-increasing number of devices necessitated the rollout of the 128-bit IPv6 addressing scheme, enabling these devices to be routable over the internet. 

\section{implementation}
\label{sec:implementation}
We demonstrate the feasibility of our proposed UWB data network and additional enhancements by implementing them in patient rooms at \emph{Katholisches Klinikum Bochum, Germany}. We deployed 40 Qorvo 3001CDK UWB development boards to construct a UWB network comprising 3 \emph{LeaderNodes}, each mimicking a \emph{zone}. The CIR data~(10,000 samples/node) from the nodes is then preprocessed to eliminate bias from signal strength. Normalization and centering force all signals to share the same peak timing, focusing the model on device fingerprints. The DL model utilizes ReLU activations for non-linearity and batch normalization for training efficiency. The Adam optimizer trains the model with a learning rate of $10^{-4}$ and a batch size of 512 over 25 epochs.

\section{Conclusion}
\label{sec:conclusion}

This paper  highlighted the missing features of UWB technology, emphasizing the imperative need for its enhancements. To strengthen the security aspect of  UWB, we proposed a novel machine learning based approach to device fingerprinting utilizing CIR information for robust  device authentication. This offers an alternative to the traditional  authentication methods. The results demonstrate the effectiveness of our proposed methods in extracting and identifying unique device fingerprints across  three distinct scenarios. Furthermore, we  explored the concept of sovereign UWB data network, envisioning its potential application in healthcare and secure environments where conventional wireless technologies may be limited or prohibited. We outlined a UWB mesh network architecture that encompasses  a wide range of devices based on their capability and availability, and developed a TDMA based MAC protocol, keeping in mind the limited payload capacity of  UWB's physical layer. 
Moreover, we have investigated the potential of the secrecy map to further improve the security of UWB systems. Finally, we advocated that the integration of the proposed enhancements into UWB standard would significantly contribute to strengthening the security posture of UWB and foster its wider adoption across various applications. 

\begin{acks}
    % This work is part of \textit{``\href{https://www.forschung-it-sicherheit-kommunikationssysteme.de/projekte/ultrasec}{Sicherheitsarchitektur für eine UWB-basierte IoT-Anwendungsplattform (UltraSec)"}} project, funded by the German Federal Ministry of Education and Research with the contract number  16KIS1682.

    This work is supported by the \grantsponsor{501100002347}{German Federal Ministry of Education and Research}{https://www.forschung-it-sicherheit-kommunikationssysteme.de/projekte/ultrasec} under Grant
No.:~\grantnum{501100002347}{16KIS1682}.
\end{acks}
\bibliographystyle{ACM-Reference-Format}
\bibliography{ultrasec-NetAISys}

% %%
% %% If your work has an appendix, this is the place to put it.
% \appendix









\end{document}